\documentclass[ALICE,manyauthors]{cernphprep}
\usepackage[comma,square,numbers,sort&compress]{natbib}
\usepackage{hyperref}
\usepackage{lineno}
\usepackage{xspace}
\usepackage{color}
\usepackage{bm}
\usepackage{setspace} 

\usepackage[T1]{fontenc}
\usepackage{orcidlink}
\begin{document}
%

\newcommand{\pp}           {pp\xspace}
\newcommand{\ppbar}        {\mbox{$\mathrm {p\overline{p}}$}\xspace}
\newcommand{\XeXe}         {\mbox{Xe--Xe}\xspace}
\newcommand{\PbPb}         {\mbox{Pb--Pb}\xspace}
\newcommand{\pA}           {\mbox{pA}\xspace}
\newcommand{\pPb}          {\mbox{p--Pb}\xspace}
\newcommand{\AuAu}         {\mbox{Au--Au}\xspace}
\newcommand{\dAu}          {\mbox{d--Au}\xspace}

\newcommand{\s}            {\ensuremath{\sqrt{s}}\xspace}
\newcommand{\snn}          {\ensuremath{\sqrt{s_{\mathrm{NN}}}}\xspace}
\newcommand{\pt}           {\ensuremath{p_{\rm T}}\xspace}
\newcommand{\meanpt}       {$\langle p_{\mathrm{T}}\rangle$\xspace}
\newcommand{\ycms}         {\ensuremath{y_{\rm CMS}}\xspace}
\newcommand{\ylab}         {\ensuremath{y_{\rm lab}}\xspace}
\newcommand{\etarange}[1]  {\mbox{$\left | \eta \right |~<~#1$}}
\newcommand{\yrange}[1]    {\mbox{$\left | y \right |~<~#1$}}
\newcommand{\dndy}         {\ensuremath{\mathrm{d}N_\mathrm{ch}/\mathrm{d}y}\xspace}
\newcommand{\dndeta}       {\ensuremath{\mathrm{d}N_\mathrm{ch}/\mathrm{d}\eta}\xspace}
\newcommand{\avdndeta}     {\ensuremath{\langle\dndeta\rangle}\xspace}
\newcommand{\dNdy}         {\ensuremath{\mathrm{d}N_\mathrm{ch}/\mathrm{d}y}\xspace}
\newcommand{\Npart}        {\ensuremath{N_\mathrm{part}}\xspace}
\newcommand{\Ncoll}        {\ensuremath{N_\mathrm{coll}}\xspace}
\newcommand{\dEdx}         {\ensuremath{\textrm{d}E/\textrm{d}x}\xspace}
\newcommand{\RpPb}         {\ensuremath{R_{\rm pPb}}\xspace}

\newcommand{\nineH}        {$\sqrt{s}~=~0.9$~Te\kern-.1emV\xspace}
\newcommand{\seven}        {$\sqrt{s}~=~7$~Te\kern-.1emV\xspace}
\newcommand{\twoH}         {$\sqrt{s}~=~0.2$~Te\kern-.1emV\xspace}
\newcommand{\twosevensix}  {$\sqrt{s}~=~2.76$~Te\kern-.1emV\xspace}
\newcommand{\five}         {$\sqrt{s}~=~5.02$~Te\kern-.1emV\xspace}
\newcommand{\twosevensixnn}{$\sqrt{s_{\mathrm{NN}}}~=~2.76$~Te\kern-.1emV\xspace}
\newcommand{\fivenn}       {$\sqrt{s_{\mathrm{NN}}}~=~5.02$~Te\kern-.1emV\xspace}
\newcommand{\LT}           {L{\'e}vy-Tsallis\xspace}
\newcommand{\GeVc}         {Ge\kern-.1emV/$c$\xspace}
\newcommand{\MeVc}         {Me\kern-.1emV/$c$\xspace}
\newcommand{\TeV}          {Te\kern-.1emV\xspace}
\newcommand{\GeV}          {Ge\kern-.1emV\xspace}
\newcommand{\MeV}          {Me\kern-.1emV\xspace}
\newcommand{\GeVmass}      {Ge\kern-.2emV/$c^2$\xspace}
\newcommand{\MeVmass}      {Me\kern-.2emV/$c^2$\xspace}
\newcommand{\lumi}         {\ensuremath{\mathcal{L}}\xspace}

\newcommand{\ITS}          {\rm{ITS}\xspace}
\newcommand{\TOF}          {\rm{TOF}\xspace}
\newcommand{\ZDC}          {\rm{ZDC}\xspace}
\newcommand{\ZDCs}         {\rm{ZDCs}\xspace}
\newcommand{\ZNA}          {\rm{ZNA}\xspace}
\newcommand{\ZNC}          {\rm{ZNC}\xspace}
\newcommand{\SPD}          {\rm{SPD}\xspace}
\newcommand{\SDD}          {\rm{SDD}\xspace}
\newcommand{\SSD}          {\rm{SSD}\xspace}
\newcommand{\TPC}          {\rm{TPC}\xspace}
\newcommand{\TRD}          {\rm{TRD}\xspace}
\newcommand{\VZERO}        {\rm{V0}\xspace}
\newcommand{\VZEROA}       {\rm{V0A}\xspace}
\newcommand{\VZEROC}       {\rm{V0C}\xspace}
\newcommand{\Vdecay} 	   {\ensuremath{V^{0}}\xspace}

\newcommand{\ee}           {\ensuremath{e^{+}e^{-}}} 
\newcommand{\pip}          {\ensuremath{\pi^{+}}\xspace}
\newcommand{\pim}          {\ensuremath{\pi^{-}}\xspace}
\newcommand{\kap}          {\ensuremath{\rm{K}^{+}}\xspace}
\newcommand{\kam}          {\ensuremath{\rm{K}^{-}}\xspace}
\newcommand{\pbar}         {\ensuremath{\rm\overline{p}}\xspace}
\newcommand{\kzero}        {\ensuremath{{\rm K}^{0}_{\rm{S}}}\xspace}
\newcommand{\lmb}          {\ensuremath{\Lambda}\xspace}
\newcommand{\almb}         {\ensuremath{\overline{\Lambda}}\xspace}
\newcommand{\Om}           {\ensuremath{\Omega^-}\xspace}
\newcommand{\Mo}           {\ensuremath{\overline{\Omega}^+}\xspace}
\newcommand{\X}            {\ensuremath{\Xi^-}\xspace}
\newcommand{\Ix}           {\ensuremath{\overline{\Xi}^+}\xspace}
\newcommand{\Xis}          {\ensuremath{\Xi^{\pm}}\xspace}
\newcommand{\Oms}          {\ensuremath{\Omega^{\pm}}\xspace}
\newcommand{\degree}       {\ensuremath{^{\rm o}}\xspace}

\definecolor{dgreen}{cmyk}{1.,0.,1.,0.4} 
\definecolor{orange}{cmyk}{0.,0.353,1.,0.} 
\newcommand{\orange}[1]{\textcolor{orange}{#1}}
\newcommand{\blue}[1]{{\color{blue}{#1}}}
\newcommand{\green}[1]{{\color{green}{#1}}}
\newcommand{\dgreen}[1]{{\textcolor{dgreen}{#1}}}
\newcommand{\red}[1]{{\color{red}{#1}}}
\newcommand{\magenta}[1]{{\color{magenta}{#1}}}
\def \new {\blue}
\def \old {\orange}
\def \ask {\magenta}
\def \tmp {\green}

\newcommand{\Epsilon}{\mathcal{E}}

\newcommand{\TwoFour}{\langle\cos\left[4\left(\Psi_4-\Psi_2\right)\right]\rangle_\text{GE}}
\newcommand{\TwoThree}{\langle\cos\left[6\left(\Psi_2-\Psi_3\right)\right]\rangle_\text{GE}}
\newcommand{\TwoSix}{\langle\cos\left[6\left(\Psi_6-\Psi_2\right)\right]\rangle_\text{GE}}
\newcommand{\ThreeSix}{\langle\cos\left[6\left(\Psi_6-\Psi_3\right)\right]\rangle_\text{GE}}
\newcommand{\TwoThreeFive}{\langle\cos\left[2\Psi_2+3\Psi_3-5\Psi_5\right]\rangle_\text{GE}}
\newcommand{\SecondTwoThreeFive}{\langle\cos\left[8\Psi_2-3\Psi_3-5\Psi_5\right]\rangle_\text{GE}}
\newcommand{\TwoFourSix}{\langle\cos\left[2\Psi_2+4\Psi_4-6\Psi_6\right]\rangle_\text{GE}}
\newcommand{\TwoThreeFour}{\langle\cos\left[2\Psi_2-6\Psi_3+4\Psi_4\right]\rangle_\text{GE}}
\newcommand{\TwoThreeFourFive}{\langle\cos\left[2\Psi_2-3\Psi_3-4\Psi_4+5\Psi_5\right]\rangle_\text{GE}}
\newcommand{\trento}{T$_\text{R}$ENTo}

\newcommand{\SPCTwoFour}{\langle\cos\left[4\left(\Psi_4-\Psi_2\right)\right]\rangle}
\newcommand{\SPCTwoThree}{\langle\cos\left[6\left(\Psi_2-\Psi_3\right)\right]\rangle}
\newcommand{\SPCTwoSix}{\langle\cos\left[6\left(\Psi_6-\Psi_2\right)\right]\rangle}
\newcommand{\SPCThreeSix}{\langle\cos\left[6\left(\Psi_6-\Psi_3\right)\right]\rangle}
\newcommand{\SPCTwoThreeFive}{\langle\cos\left[2\Psi_2+3\Psi_3-5\Psi_5\right]\rangle}
\newcommand{\SPCSecondTwoThreeFive}{\langle\cos\left[8\Psi_2-3\Psi_3-5\Psi_5\right]\rangle}
\newcommand{\SPCTwoFourSix}{\langle\cos\left[2\Psi_2+4\Psi_4-6\Psi_6\right]\rangle}
\newcommand{\SPCTwoThreeFour}{\langle\cos\left[2\Psi_2-6\Psi_3+4\Psi_4\right]\rangle}
\newcommand{\SPCTwoThreeFourFive}{\langle\cos\left[2\Psi_2-3\Psi_3-4\Psi_4+5\Psi_5\right]\rangle}
\begin{titlepage}
\PHyear{2023}       
\PHnumber{012}      
\PHdate{01 February}  

\title{Symmetry plane correlations in Pb--Pb collisions at $\bm{\sqrt{s_\text{NN}}}$~=~2.76~TeV } 
\ShortTitle{Symmetry plane correlations in Pb--Pb collisions at $\sqrt{s_\text{NN}}$~=~2.76~TeV }   

\Collaboration{ALICE Collaboration\thanks{See Appendix~\ref{app:collab} for the list of collaboration members}}
\ShortAuthor{ALICE Collaboration} 

\begin{abstract}
A newly developed observable for correlations between symmetry planes, which characterize the direction of the anisotropic emission of produced particles, is measured in Pb--Pb collisions at $\sqrt{s_\text{NN}}$~=~2.76~TeV with ALICE. This so-called Gaussian Estimator allows for the first time the study of these quantities without the influence of correlations between different flow amplitudes. The centrality dependence of various correlations between two, three and four symmetry planes is presented. The ordering of magnitude between these symmetry plane correlations is discussed and the results of the Gaussian Estimator are compared with measurements of previously used estimators. The results utilizing the new estimator lead to significantly smaller correlations than reported by studies using the Scalar Product method. Furthermore, the obtained symmetry plane correlations are compared to state-of-the-art hydrodynamic model calculations for the evolution of heavy-ion collisions. While the model predictions provide a qualitative description of the data, quantitative agreement is not always observed, particularly for correlators with significant non-linear response of the medium to initial state anisotropies of the collision system. As these results provide unique and independent information, their usage in future Bayesian analysis can further constrain our knowledge on the properties of the QCD matter produced in ultrarelativistic heavy-ion collisions.

\end{abstract}
\end{titlepage}

\setcounter{page}{2} 


\section{Introduction} 

One of the most important discoveries in the physics of heavy-ion collisions at ultrarelativistic energies is the observation of a deconfined state of nuclear matter dubbed quark--gluon plasma (QGP). This extreme state is produced during the heavy-ion collision evolution, and its properties resemble the properties of a perfect liquid. Unprecedentedly large data sets collected at the LHC enable the most quantitative description of the QGP to date. Given the complexity of the system produced in heavy-ion collisions, an important program in the field is the development of observables that provide new and independent information inaccessible with previous measurements~\cite{Gale:2013da,Heinz:2013th,Braun-Munzinger:2015hba,Bernhard:2016tnd,Pasechnik:2016wkt,Busza:2018rrf,Bernhard:2019bmu,ALICE:2022wpn}.

The intersecting volume of two heavy ions is anisotropic in coordinate space, either due to collision geometry (particularly in non-central collisions with large values of impact parameter) or due to fluctuations of positions of participating nucleons (most significant in central head-on collisions). Anisotropic pressure gradients, which develop in this volume containing the strongly interacting nuclear matter, transfer the initial-state spatial anisotropies into final-state anisotropies in momentum space. This phenomenon is known as anisotropic flow and it is a sensitive probe of all stages in the heavy-ion collision evolution~\cite{Ollitrault:1992bk}. Anisotropic flow measurements are used to constrain the transport properties of the QGP, for instance ratios of shear and bulk viscosities to entropy density
~\cite{Bernhard:2016tnd,Bernhard:2019bmu,Nijs:2020roc,JETSCAPE:2020mzn,Parkkila:2021tqq,Parkkila:2021yha}. The anisotropic emission of particles in the plane transverse to the beam direction is quantified with amplitudes $v_n$ and symmetry planes $\Psi_n$ by using the Fourier series decomposition  of the azimuthal angle ($\varphi$) distribution of produced particles~\cite{Voloshin:1994mz}
\begin{equation}
f(\varphi)=\frac{1}{2\pi}\left[1+2\sum_{n=1}^{\infty}v_n\cos[n(\varphi-\Psi_n)]\right]\,.
\label{eq:FourierSeries}
\end{equation}
A detailed discussion of properties of $v_n$ and $\Psi_n$ can be found in Ref.~\cite{Bilandzic:2020csw}. The symmetry plane $\Psi_n$ has a simple geometrical interpretation when the anisotropic distribution can be parameterized only with one harmonic $n$, since then it can be shown that 
$f(\Psi_n+\varphi) = f(\Psi_n-\varphi)$, i.e. symmetry plane~$\Psi_n$ is the plane for which it is equally probable for a particle to be emitted above or below it.

Historically, the emphasis was on studying the amplitudes $v_n$, but the symmetry planes also carry a very important information about different stages in heavy-ion collision evolution. Unlike the flow amplitudes $v_n$, a single symmetry plane $\Psi_n$ cannot be estimated directly in an experiment using correlation techniques --- the simplest available observables are symmetry plane correlations (SPC), for instance $\langle\cos 4(\Psi_4-\Psi_2)\rangle$~\cite{Jia:2012ma,Bilandzic:2020csw}. Such correlations are the subject of this study.

In the early anisotropic flow analyses, the goal was to measure $v_n$ with respect to the reaction plane (a plane spanned by the beam axis and impact parameter vector), and it was assumed that all symmetry planes are approximately the same and equal to the orientation of the reaction plane. Therefore, the first flow measurements were exclusively of flow amplitudes $v_n$. The first experimental results for SPC can be traced back to the E877 experiment~\cite{E877:1996czs}. These initial measurements were  performed by the standard event plane method with the subevent technique~\cite{Poskanzer:1998yz}. The first measurements of SPC involving two symmetry planes in the RHIC era were obtained by PHENIX in Refs.~\cite{PHENIX:2009cjr,PHENIX:2011yyh}. An alternative approach was pursued by NA49 and STAR using 3-particle mixed-harmonic correlations, which by definition have contributions from SPC~\cite{NA49:2003njx,STAR:2003xyj}. In the first flow studies at LHC energies, the ALICE Collaboration demonstrated in Ref.~\cite{ALICE:2011ab} that the symmetry planes $\Psi_2$ and $\Psi_3$ fluctuate independently in all considered centralities. Finally, the most detailed experimental results to date were published by the ATLAS Collaboration in Ref.~\cite{ATLAS:2014ndd}, where also for the first time the strength of correlations among three symmetry planes was presented. ATLAS systematically studied the centrality dependence of SPC both in the initial and final state using the analysis technique from Refs.~\cite{Jia:2012ma,Jia:2012ju}. It was concluded that SPC originate both from correlated fluctuations in the initial geometry and from the non-linear mixing between different flow harmonics in the final state. Subsequent experimental publications which used SPC to constrain the details of the non-linear hydrodynamic response can be found in Refs.~\cite{ATLAS:2015qwl,ALICE:2017fcd,CMS:2019nct,ALICE:2020sup,STAR:2022vqw}.

In theoretical studies, SPC can be obtained directly both in coordinate and in momentum space~\cite{Miller:2007ri,Alver:2010gr,Staig:2010pn,Nagle:2010zk,Jia:2012ma,Jia:2012ju,Qin:2011uw,Qiu:2012uy,ATLAS:2014ndd,Yan:2015fva,Teaney:2010vd}. State-of-the-art modeling of heavy-ion collisions covers all stages of its evolution starting from the initial conditions to the final free streaming of produced particles. The SPC in the initial state can be obtained event-by-event directly from the underlying model of the collision geometry using for instance energy density distribution or nucleon positions, while in the final state SPC are the event-by-event output of the model used to describe all subsequent stages in the evolution. Therefore, in theoretical studies it is not, in general, necessary to build an estimator for SPC from the azimuthal angles of final-state particles, like it is done in an experiment. In order to ease the comparison between theoretical and experimental results, azimuthal correlators were used to indirectly estimate SPC also in Refs.~\cite{Bhalerao:2011yg,Bhalerao:2011ry,Bhalerao:2011bp,Luzum:2012da,Bhalerao:2013ina,Niemi:2015qia,Nijs:2020roc}. Other types of theoretical studies involving symmetry planes can be found in Refs.~\cite{Petersen:2010cw,Qin:2010pf,Lacey:2010av,Qiu:2011iv,Heinz:2013bua,Xu:2020koy}.  

Several experimental difficulties associated with the SPC render their measurements particularly challenging. Even in the simplest realisation, it is necessary to construct non-trivial estimators for SPC to resolve these issues. Unlike the flow amplitudes $v_n$, each symmetry plane $\Psi_n$ taken individually is not invariant under rotations of the coordinate system in the laboratory frame in which azimuthal angles are measured (see Eq.~\ref{eq:FourierSeries}). Therefore, the simplest rotationally-invariant physical observable involving symmetry planes is the difference of two symmetry planes. In an actual experiment such rotations are unavoidable as a direct consequence of random event-by-event fluctuations of the direction of the impact parameter vector. Only symmetry planes that are different, apart for trivial periodicity, carry independent information, and therefore any dependence on periodicity must be removed from all SPC observables by definition. The widely used technique to suppress systematic biases from short-range nonflow correlations by introducing pseudorapidity gaps in the measured azimuthal correlators which are used to estimate SPC is not applicable due to decorrelations of symmetry planes as a function of pseudorapidity~\cite{ATLAS:2017rij,Bozek:2017qir,ATLAS:2020sgl,Sakai:2020pjw,Xu:2020koy,Cimerman:2021gwf,Sakai:2021rug}. Moreover, it has been shown recently that the effect of flow magnitude correlations, which have been either completely~\cite{ATLAS:2014ndd} or partially~\cite{ALICE:2017fcd} neglected in the existing measurements, may overshadow the correlations of symmetry planes in the analysis with the Scalar Product~(SP) method~\cite{Bilandzic:2020csw}. The new and improved estimator for SPC from Ref.~\cite{Bilandzic:2020csw}, which overcomes these limitations, is introduced next.
    
The starting point is the following relation between $v_n$ and $\Psi_n$, and multiparticle azimuthal correlations~\cite{Bhalerao:2011yg,Bilandzic:2013kga,Bilandzic:2020csw}:
\begin{equation}
v_{n_1}^{a_1}\cdots v_{n_k}^{a_k}\,e^{i(a_1 n_1\Psi_{n_1}+\cdots+a_k n_k\Psi_{n_k})} = \left<e^{i(n_1\varphi_1+\cdots+n_l\varphi_l)}\right> \,.
\label{eq:generalResult}
\end{equation}
In this equation, angular brackets indicate an average over the azimuthal angles of all distinct sets of $l$ particles measured in the same event.

The coefficients $a_i$ are positive integers which ensure that all harmonics $n_i$ and symmetry planes $\Psi_{n_i}$ are unique on the left-hand side in the above expression. These coefficients can be understood in the following way: $a_i$ counts how many times a harmonic $n_i$ appears in the azimuthal correlator on the right-hand side of Eq.~(\ref{eq:generalResult}) (harmonics with positive and negative signs are counted separately). The total number of particles, i.e. the order of the multiparticle azimuthal correlator, is given by $\sum_i a_i$. The index $k$ on the left-hand side labels only unique harmonics in the original set $n_1, n_2, \ldots, n_l$, therefore $k\leq l$. As an example, for the correlator $\left<e^{i(2\varphi_1\!-\!\varphi_2\!-\!\varphi_3)}\right>$ it follows that $n_1=2, a_1=1, n_2=n_3=-1, a_2=2$. The advantage of this generalized notation is that now $n_i$ and $a_i$ decouple naturally either into a subscript or into an exponent when associated with flow amplitudes $v_{n_i}$ in Eq.~(\ref{eq:generalResult}), which enables their distinct physical interpretation. Finally, solely from the definition of the Fourier series in Eq.~(\ref{eq:FourierSeries}) one can prove that $v_{-n} = v_n$ and $\Psi_{-n} = \Psi_n$, which is used in the rest of the paper. Due to this property, the final $a$ coefficient for harmonic $n$ in Eq.~(\ref{eq:generalResult}) is a sum $a_n + a_{-n}$.

Taking into account all these technical considerations, the simplest definition of SPC observables is provided by the following expression~\cite{Bhalerao:2011yg,Jia:2012ju,ATLAS:2014ndd}: 
\begin{equation}
\left<e^{i(a_1n_1\Psi_{n_1} + \cdots + a_kn_k\Psi_{n_k})}\right>,\quad\sum_i^k a_i n_i = 0\,,
\label{eq:symmetry-plane-observables}
\end{equation}
where all $a_i$ are positive and all $n_i$ are unique integers. Angular brackets $\langle\ \rangle$ indicate here an average over all events. Defined this way, SPC observables are rotationally invariant and therefore invariant with respect to random event-by-event fluctuations of the impact parameter vector, while the periodicity of each individual symmetry plane is accounted for by definition. Experimentally, Eq.~(\ref{eq:generalResult}) is used as a starting point for an estimator for SPC. However, to isolate the true SPC part, the prefactor $v_{n_1}^{a_1}\cdots v_{n_k}^{a_k}$ has to be divided out. The importance of this technical detail was neglected in all previously used SPC estimators.

The new and improved SPC estimator, named the {\it Gaussian Estimator} (GE), was developed recently in Ref.~\cite{Bilandzic:2020csw}. Its key improvement amounts to using the following expression to estimate SPC:
\begin{equation} \label{GaussApproxExp}
\begin{split}
&	\langle \cos\left(a_{1} n_1 \Psi_{n_1}+\cdots+ a_{k} n_k \Psi_{n_k}\right) \rangle_{\text{GE}} =  \sqrt{\frac{\pi}{4}}\;\;\;\frac{\langle  v_{n_1}^{a_{1}} \;\cdots\; v_{n_k}^{a_{k}}\cos\left(a_{1} n_1 \Psi_{n_1}+\cdots+a_{k} n_k \Psi_{n_k}\right) \rangle}{\sqrt{\langle  v_{n_1}^{2a_{1}} \;\cdots\;v_{n_k}^{2a_{k}}  \rangle}},
\end{split}
\end{equation}
which was derived by approximating multi-harmonic flow fluctuations with a two-dimensional Gaussian distribution. Both the numerator and denominator on the right-hand side in the above expression can be estimated by using Eq.~(\ref{eq:generalResult}) with suitably chosen harmonics $n_i$. Further explanations of the technical details of the GE based on the example $\SPCTwoFour$ are provided in Appendix~\ref{Sec. GE-Example}. The main improvement of this new estimator can be found in the denominator where the GE has the joined multivariate moment of different flow amplitudes, $\langle v_{n_1}^{2a_{1}} \;\cdots\;v_{n_k}^{2a_{k}}\rangle$. This is in contrast to the previously used SP estimator, defined as~\cite{Bhalerao:2013ina}
\begin{equation}
\langle \cos\left(a_{1} n_1 \Psi_{n_1}+\cdots+a_{k} n_k \Psi_{n_k}\right) \rangle_{\text{SP}}=\frac{\langle v_{n_1}^{a_{1}} \;\cdots\; v_{n_k}^{a_{k}} \cos\left(a_{1} n_1 \Psi_{n_1}+\cdots+a_{k} n_k \Psi_{n_k}\right)\rangle}{\sqrt{\langle v_{n_1}^{2a_{1}} \rangle \cdots \langle v_{n_k}^{2a_{k}} \rangle}}\,,
\label{eq:SP}
\end{equation}
which uses instead $\langle  v_{n_1}^{2a_{1}}\rangle \;\cdots\;\langle v_{n_k}^{2a_{k}} \rangle$ in the denominator and therefore assumes that event-by-event fluctuations of flow amplitudes are mutually independent. This assumption is in contradiction with recent experimental results which reported strong and non-trivial correlated fluctuations of different flow amplitudes, both at RHIC and LHC energies, and across different collisions systems~\cite{ATLAS:2015qwl,ALICE:2016kpq,Acharya:2017gsw,CMS:2017kcs, STAR:2018fpo}. These shortcomings of the previous SPC results are the main motivation for the current work. As it was pointed out in Ref.~\cite{Bilandzic:2020csw}, the GE does not account for cross-correlations between flow amplitudes and symmetry planes. However, the study in Ref.~\cite{Bilandzic:2020csw} showed that the contribution by these cross-correlations is minor when compared to the correlations between flow amplitudes. 

The rest of the article is organized as follows. In Sec.~\ref{s:DataAnalysis} the ALICE detector is introduced, together with the analyzed data set and analysis details, such as the event and track selection criteria. In Sec.~\ref{s:Results} the SPC results using the GE are presented, comparisons with previous experimental results are discussed, and confrontation with state-of-the-art theoretical models is displayed. The article concludes in Sec.~\ref{s:Conclusion} with the summary. A more detailed discussion about the technical details of the GE can be found Appendix~\ref{Sec. GE-Example}.

\section{Data Analysis} 
\label{s:DataAnalysis}
The data set consists of Pb--Pb collisions at a center-of-mass energy per nucleon pair $\sqrt{s_{\text{NN}}}=2.76~\text{TeV}$ recorded by ALICE in 2010. A detailed description of the apparatus and its performance is given in Refs.~\cite{ALICE:2008ngc,ALICE:2014sbx}. The Silicon Pixel Detector (SPD), which comprises the two innermost layers of the Inner Tracking System (ITS)~\cite{ALICE:1999cls,ALICE:2010tia}, and both V0 detectors ~\cite{ALICE:2013axi} were used for triggering. The latter consists of two arrays of scintillator counters, the V0A and V0C, covering a pseudorapidity range of $2.8<\eta<5.1$ and $-3.7<\eta<-1.7$, respectively. The SPD covers pseudorapidities of $|\eta| < 2.0$ for its inner and $|\eta| < 1.4$ for its outer layer. Minimum bias collisions were selected by requiring a signal in at least two out of the three following: two chips in the outer layer of the SPD, the V0A, and the V0C. 
For this analysis, only events with a primary vertex within $\pm10~\text{cm}$ of the nominal interaction point along the beam axis were used. The centrality of the collisions~\cite{ALICE:2013hur} was estimated with the SPD. Backgrounds events due to beam--gas interactions and parasitic beam--beam interactions were removed by using V0 and Zero Degree Calorimeter~\cite{ALICE:1999edx} timing information. Overall, after the event selection the used data set consists of $7.36\times10^6$ reconstructed collisions for the centrality range 0--50\%.

The reconstruction of charged particle trajectories was performed using only information from the Time Projection Chamber (TPC)~\cite{Alme:2010ke} due to its uniform acceptance in azimuth. This analysis used tracks with transverse momenta $0.2<p_{\text{T}}<5.0~\text{GeV}/c$ and in a pseudorapidity range of $|\eta|<0.8$, while covering the full azimuth. The lower boundary of the transverse momentum selection ensured a large and stable tracking efficiency in the TPC, while the upper cutoff decreases the contribution from jets which in general have larger momenta. The charged tracks were accepted for the analysis if they had a minimum of 70 out of a maximum of 159 space points in the TPC. The $\chi^2$ per space point from the track fit was set to be within $0.1<\chi^2/ {\rm NDF}<4.0$. The distance of closest approach (DCA) of the extrapolated tracks to the primary vertex was required to be at maximum 2.4~cm in the transverse plane and 3.2~cm in the beam direction. Daughter tracks with a reconstructed secondary weak-decay kink topology (i.e. tracks with an abrupt change of direction) were discarded. The contamination from secondaries as well as the reconstruction efficiency with this track selection can be found in Ref.~\cite{ALICE:2021klf}. 

The $p_{\text{T}}$-dependent reconstruction efficiency was corrected using particle weights according to Ref.~\cite{Bilandzic:2013kga}. These weights were obtained with the HIJING (Heavy-Ion Jet INteraction Generator) Monte Carlo generator~\cite{Gyulassy:1994ew} by comparison of generated and reconstructed tracks. For the latter, a GEANT3~\cite{Brun:1082634} detector simulation and event reconstruction was used in addition to HIJING. At the same time, weights to correct for non-uniform acceptance in azimuthal angle did not have to be applied due to the uniform acceptance of the TPC over the whole azimuth in the analyzed data set. Nonflow contributions, i.e. correlations between a few particles unrelated to collective anisotropic flow, were investigated with HIJING for the numerator and denominator of the GE in Eq.~\eqref{GaussApproxExp} separately. For all SPC combinations, both the numerator and denominator were found to be consistent with zero in all considered centrality ranges, demonstrating that the analyzed SPC observables are not influenced by most important sources of nonflow correlations such as jets or resonance decays.

The statistical uncertainties of the measured SPC were obtained via propagation of uncertainties of the numerator and denominator in Eq.~\eqref{GaussApproxExp}. Systematic uncertainties were evaluated by varying the default event and track selections. All variations were performed one at a time and only those with a difference larger than $2\sigma$, where $\sigma$ is the uncertainty of the difference, with respect to the default selection were taken into account for the final systematic uncertainty. All individual systematic variations were considered independent and combined in quadrature to obtain the total systematic uncertainty. Regarding the event selection criteria, the position of the primary vertex along the beam line was varied to $\pm6~\text{cm}$ and $\pm8~\text{cm}$, where a relative effect on the measured observables of up to 5\% was found. A systematic uncertainty of up to 6\% from the centrality estimation was determined by using the V0 instead of the SPD. To evaluate the uncertainty due to the track selection, the number of TPC clusters used in the track reconstruction was varied to a required minimum of 80, 90 and 100 compared to the default 70. This resulted in a systematic uncertainty of up to 6\%. The sensitivity of the results to the track quality was checked by varying the $\chi^2/ {\rm NDF}$ to $0.3<\chi^2/ {\rm NDF}<4.0$ and $0.1<\chi^2/ {\rm NDF}<3.5$, which led to an additional uncertainty of up to 6\%.  Two variations were performed regarding the DCA by changing the upper limit in the transverse direction to 1~cm and in the longitudinal direction to 2~cm. The variation of the DCA changes the contribution from secondaries in the analysis as these particles usually have a larger DCA than primary particles. The DCA variation in the transverse plane led to a systematic uncertainty of about 3--10\%, while the check along the beam axis had a relative variation of about 4\%. Additionally, an independent analysis was performed by using a different track reconstruction procedure, which employs combined information from both the TPC and the ITS. This led to an uncertainty in the range of 5--10\%.

\section{Results}
\label{s:Results}
\begin{figure}[!htbh]
	\centering
	\includegraphics[width=\textwidth]{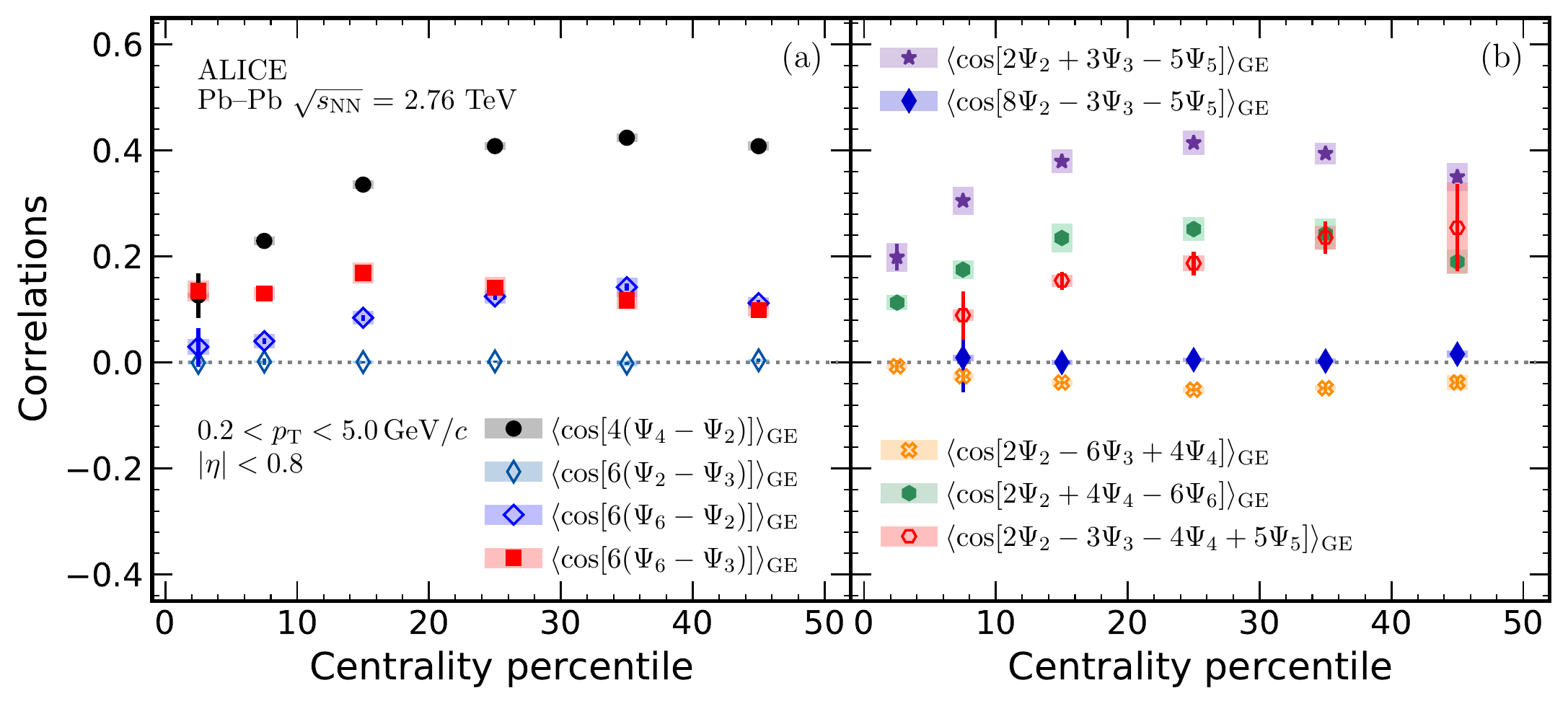}
	\caption{Comparison of the extracted correlations between different combinations of two symmetry planes (a) and between three and four planes (b) using the GE in Eq.~\eqref{GaussApproxExp}. Statistical (systematic) uncertainties are shown as lines (boxes).} 
	\label{Fig1Allobs}
\end{figure}

The centrality dependence of the correlations between different combinations of two and three symmetry planes, as well as the first measurement of a correlation between four planes, are presented in  Fig.~\ref{Fig1Allobs}. In the case of two symmetry planes shown in Fig.~\ref{Fig1Allobs}(a), the strongest correlation is observed for $\TwoFour$, while the correlation strength gets weaker for $\TwoSix$ and $\ThreeSix$. The results for $\TwoThree$ are compatible with zero within uncertainties. A hierarchy, $\TwoFour$ $>$ $\ThreeSix$ $>$  $\TwoSix$, holds for the centrality range 5--50\%, with an exception of $\ThreeSix$ and $\TwoSix$ being comparable at centralities above 20\%. The details of the centrality dependence vary for the different combinations of symmetry planes. While $\TwoFour$ and $\TwoSix$ are increasing non-linearly from central to semicentral collisions, $\ThreeSix$ shows a weak centrality dependence. The observed zero signal for $\TwoThree$ indicates that no correlation is present within the current uncertainties for the final-state planes $\Psi_2$ and $\Psi_3$, while $v_2$ and $v_3$ are anti-correlated~\cite{ALICE:2016kpq,Acharya:2017gsw,Niemi:2012aj, Jia:2014jca, ATLAS:2015qwl, Qian:2016pau, Niemi:2015qia}. This result justifies the necessity of measuring separately correlations of symmetry planes and flow magnitudes, because these measurements can be used to independently constrain properties of the matter produced in heavy-ion collisions.

 The different magnitudes of correlations are also observed for three symmetry planes as shown in Fig.~\ref{Fig1Allobs}(b). The magnitude and details of the centrality dependence vary for different combinations of flow harmonics. The $\TwoThreeFive$ observable exhibits the strongest correlations and $\TwoThreeFour$ shows the weaker signal. The SPC $\TwoThreeFour$ is the only correlator with a negative sign, which will be discussed later on in more detail.
The $\SecondTwoThreeFive$ observable is consistent with zero within uncertainties, similar to $\TwoThree$.
The correlation between four planes, $\TwoThreeFourFive$ shows the strongest centrality dependence among all harmonic combinations and increases towards peripheral collisions.

 The magnitudes of SPC are ordered approximately based on the corresponding order of the particle correlations. The two largest SPCs, $\TwoFour$ and $\TwoThreeFive$, are both measured with three-particle correlators. In contrast, the smallest ones are $\TwoThree$ and $\SecondTwoThreeFive$, which are five- and six-particles correlations, respectively. One possible explanation is the following: the flow vector fluctuations encoded in the observed correlations are mainly attributed to the fluctuation of the initial state. Also, the initial state fluctuation is attributed to the fluctuation of a finite amount of ``sources'' produced at the degrees of freedom collision points, namely protons and neutrons, in the collision region. The Central Limit Theorem (CLT) states that for independent random variables (here, the position of sources), the sample average tends toward a Gaussian distribution when the number of sampling increases. A clear example of such behavior was studied for initial ellipticity in Ref.~\cite{Yan:2014afa}, where it was shown how the ellipticity fluctuation distribution changes from elliptic-power distribution with large skewness to a Gaussian distribution at a large number of sources. The order of particle correlations corresponds to the order of the cumulants of the underlying flow vector fluctuation. To see the clear connection, correlations should be written in a Cartesian notation rather than polar notation (see Refs.~\cite{Giacalone:2016eyu,Abbasi:2017ajp,Mehrabpour:2018kjs} for the relation between skewness and Kurtosis of flow vector distribution to the particle correlations). Only the second-order cumulant, namely the width of the distribution, is nonvanishing for a Gaussian distribution. As a result, higher-order cumulants (skewness, kurtosis, etc.) are small for distributions close to Gaussian. These studies are done for flow amplitudes with only one harmonic, but the logic is true for more than one harmonic as well. The observed ordering of magnitudes in Fig.~\ref{Fig1Allobs} indicates that the contribution of higher-order cumulants is smaller compared to lower ones in general, meaning the lowest-order cumulants have the dominant role in deviation from Gaussianity. A crossing between $\ThreeSix$ (a three-particle correlation) and $\TwoSix$ (a four-particle correlation) is observed with centralities above 25\% where the number of final state particles is lower. The same is true for $\TwoFourSix$ (a three-particle correlation) and $\TwoThreeFourFive$ (a four-particle correlation). The effect of non-Gaussianity is expected to be more dominant in this centrality region since the system size is smaller and less number of sources are expected.  At a finite number of sources, the actual ordering of the correlation magnitudes depends on the details of the underlying source fluctuation that needs a separate study.

\begin{figure}[!htbh]
	\centering
	\includegraphics[width=\textwidth]{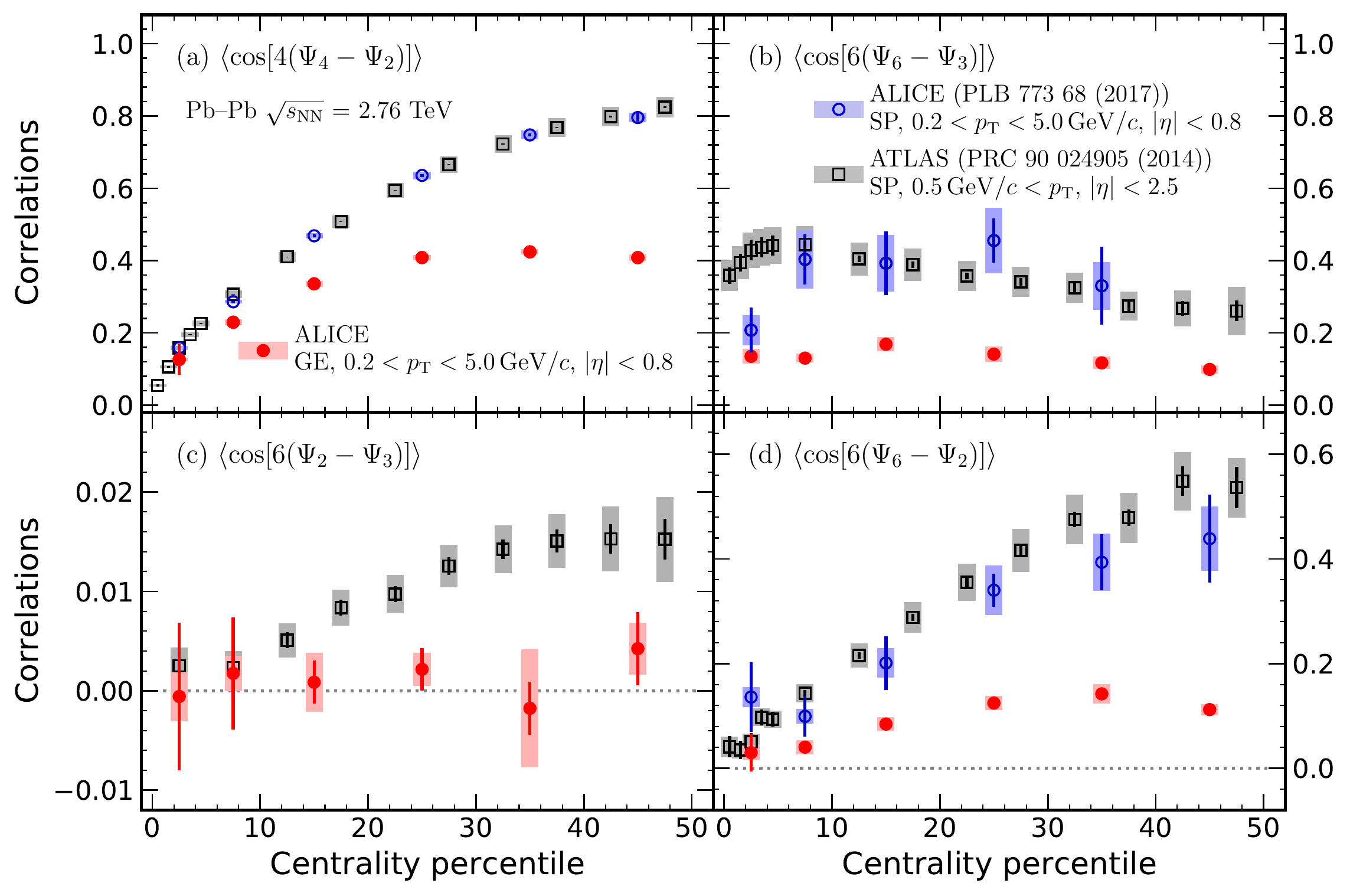}
\caption{Experimental data of correlations between two symmetry planes obtained with the GE compared with measurements from ATLAS~\cite{ATLAS:2014ndd} and ALICE~\cite{ALICE:2017fcd} using the SP method. Statistical and systematic uncertainties are represented by lines and boxes, respectively.} 
	\label{Fig4_TwoHar_SP}
\end{figure}

\begin{figure}[!htbh]
	\centering
	\includegraphics[width=\textwidth]{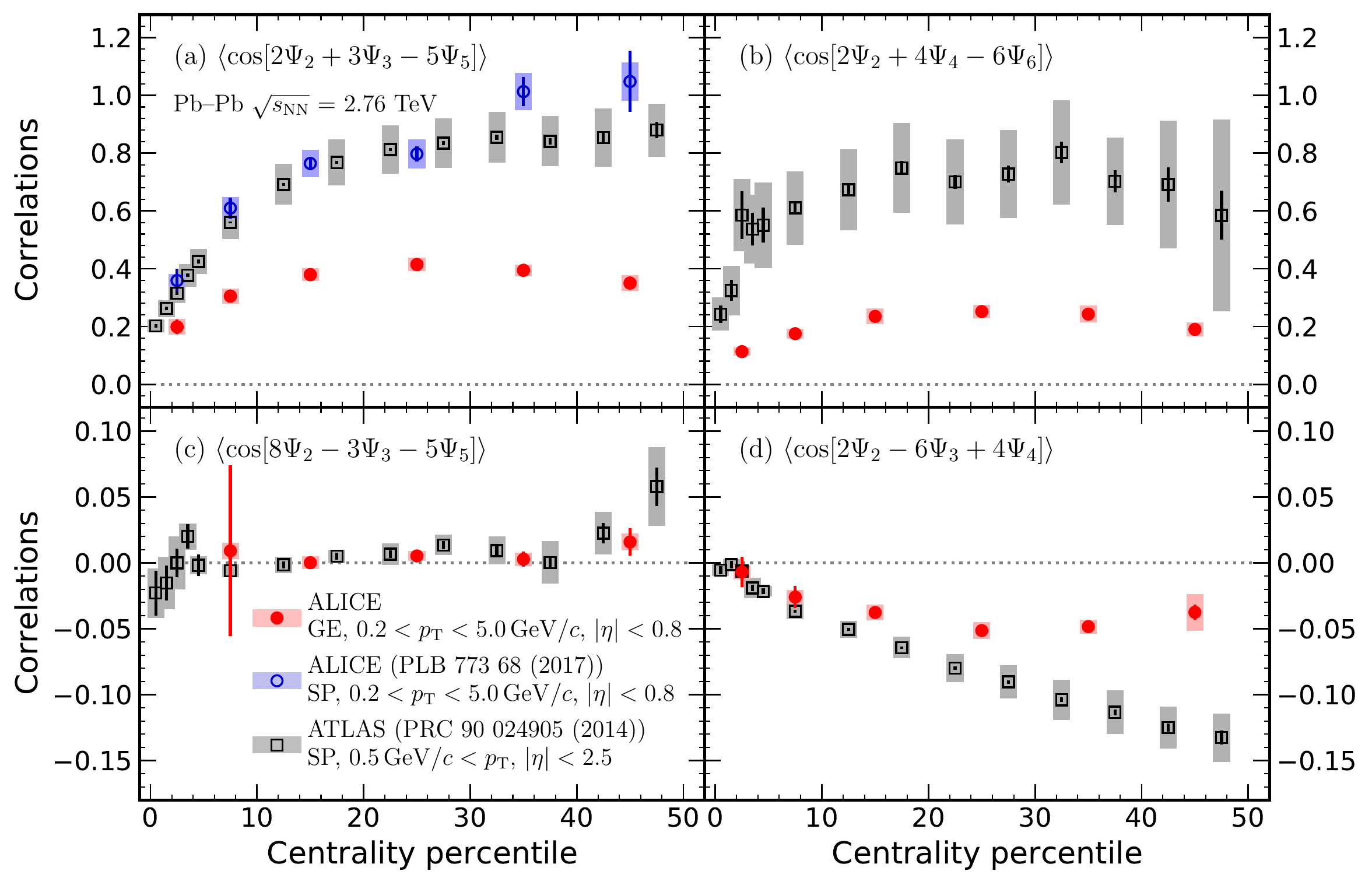}
	\caption{Correlations between three symmetry planes obtained with the GE compared with measurements from ATLAS~\cite{ATLAS:2014ndd} and ALICE~\cite{ALICE:2017fcd} using the SP method. Statistical and systematic uncertainties are shown as lines and boxes, respectively.} 
	\label{Fig5_ThreeHar_SP}
\end{figure}

In Figs.~\ref{Fig4_TwoHar_SP} and~\ref{Fig5_ThreeHar_SP} the experimental data for SPC estimated with the GE are compared with the results obtained from ATLAS~\cite{ATLAS:2014ndd} and ALICE~\cite{ALICE:2017fcd} using the SP method. While the analysis of the SP method by ALICE used the same kinematic range as the work presented in this article, the analysis by ATLAS was performed in a wider range of $0.5~\text{GeV}/c < p_\text{T}$ and $|\eta|<2.5$. Despite this difference in kinematic regions, the SPC extracted by the SP method from ALICE and ATLAS agree within uncertainties. In general, the obtained data from the GE are significantly smaller than the estimates performed with the SP method for centralities larger than 10\%. This difference is mainly attributed to the fact that correlations between flow amplitudes were not removed in the SP method as it was demonstrated in Ref.~\cite{Bilandzic:2020csw}. For the SPC $\SPCTwoFour$ and $\SPCTwoSix$ shown in Fig.~\ref{Fig4_TwoHar_SP}, the GE and SP method are compatible only in 0--5\% centrality, while for $\SPCThreeSix$ the GE differs in all centrality intervals when compared to the SP method by ATLAS. For centralities larger than 5\%, a clear splitting between all of the previously mentioned SPC is visible with significantly smaller values obtained by the GE. For $\SPCTwoThree$ the experimental data of the GE are compatible with zero within the uncertainties in all considered centrality intervals. In contrast to that, the results of the SP method show a small, but non-zero value. 
However, the results obtained with the GE show larger uncertainties when compared to the SP for this particular SPC. Future studies with larger data sets will show whether the SPC $\SPCTwoThree$ remains compatible with zero within uncertainties when using the GE or if a small non-zero correlation exists which cannot be resolved within the present uncertainties. In the latter case, the results of the GE will nonetheless lead to significantly smaller values than reported by the SP method. 

Similarly, the experimental results of the GE and the SP method are compared to each other for SPC between three planes. The results are presented in Fig.~\ref{Fig5_ThreeHar_SP}. For the combinations $\SPCTwoThreeFive$, $\SPCTwoFourSix$ and $\SPCTwoThreeFour$ the GE again leads to significantly smaller values than the SP method for centralities larger than 10\%. For $\SPCTwoThreeFive$ it has to be noted that the observables previously employed by ALICE~\cite{ALICE:2017fcd} and ATLAS~\cite{ATLAS:2014ndd} differ in the denominator. ATLAS uses a fully factorized denominator $\langle v_2^2\rangle \langle v_3^2 \rangle \langle v_5^2\rangle$ as in the definition of the SP method~\eqref{eq:SP}, while the denominator in the ALICE measurement is only partially factorized $\langle v_2^2v_3^2 \rangle \langle v_5^2\rangle$ and thus is not defined exactly as in Eq.~\eqref{eq:SP}. To ease the notation in Fig.~\ref{Fig5_ThreeHar_SP} we still label $\SPCTwoThreeFive$ measured by ALICE~\cite{ALICE:2017fcd} as SP method. The SPC $\SPCSecondTwoThreeFive$ is the only combination where the estimates by the GE and the SP method are compatible with each other within uncertainties in all considered centralities, as the results from the SP method are already close to zero. The difference in physical interpretation between the two SPC involving $\Psi_2$, $\Psi_3$ and $\Psi_5$ is discussed later.

\begin{figure}[!htbh]
	\centering
	\includegraphics[width=\textwidth]{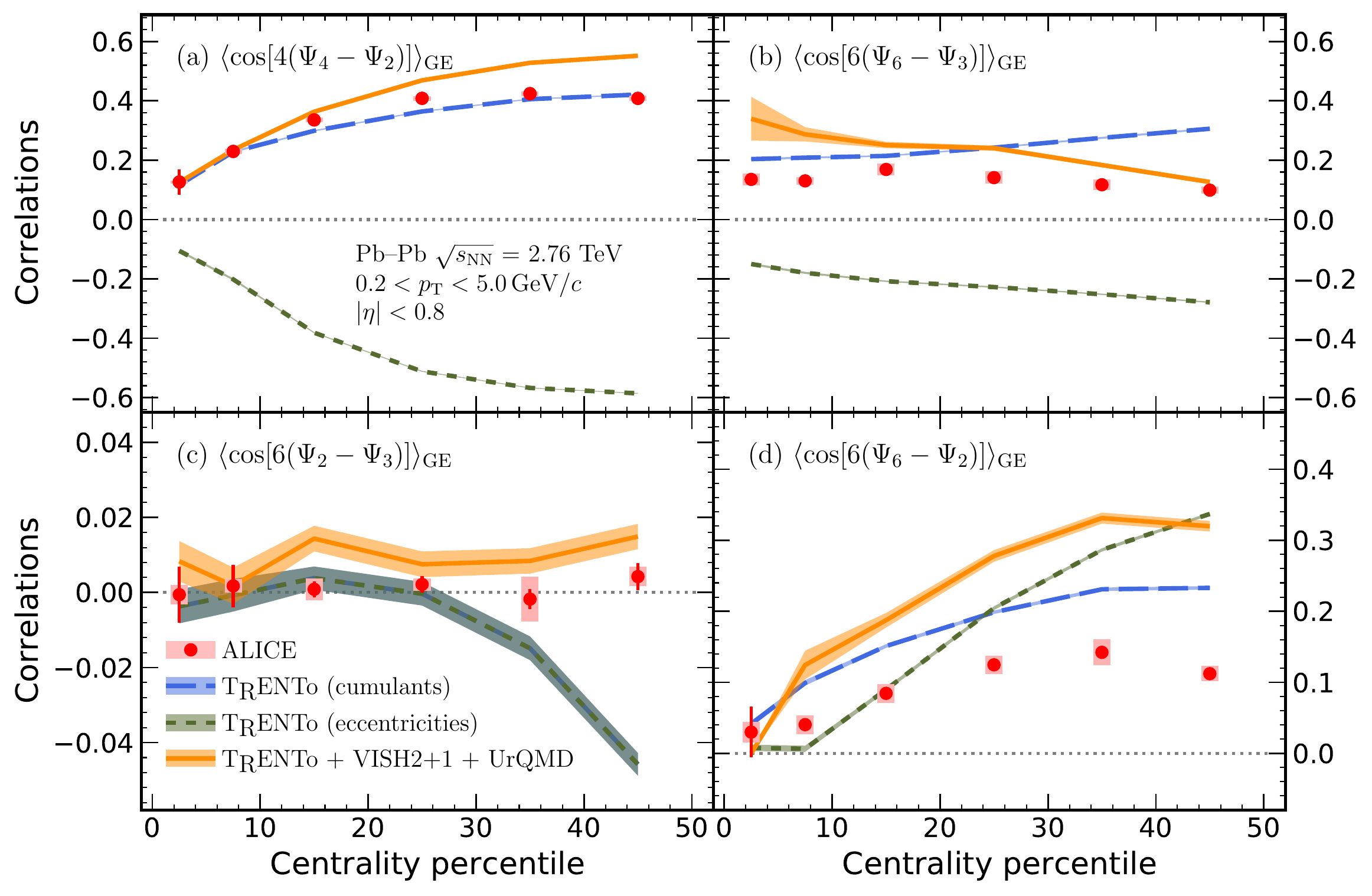}
\caption{Experimental data for correlations between two symmetry planes compared to theoretical predictions in the initial and final state obtained with \trento\:and \trento+VISH(2+1)+UrQMD~\cite{Moreland:2014oya,Shen:2014vra,Song:2007ux,Bass:1998ca,Bleicher:1999xi}, respectively. For $\TwoThree$ (c), the initial state predictions calculated via eccentricities and energy density cumulants, $\langle \cos [6(\phi_2-\phi_3)]\rangle_{\text{GE}}$ and $\langle \cos[6(\Phi_2-\Phi_3)]\rangle_{\text{GE}}$, fully overlap. Statistical (systematic) uncertainties of the ALICE data are shown as lines (boxes). The statistical uncertainties of the models are represented by the colored bands.} 
	\label{Fig2_TwoHar_models}
\end{figure}
 
\begin{figure}[!htbh]
	\centering
	\includegraphics[width=\textwidth]{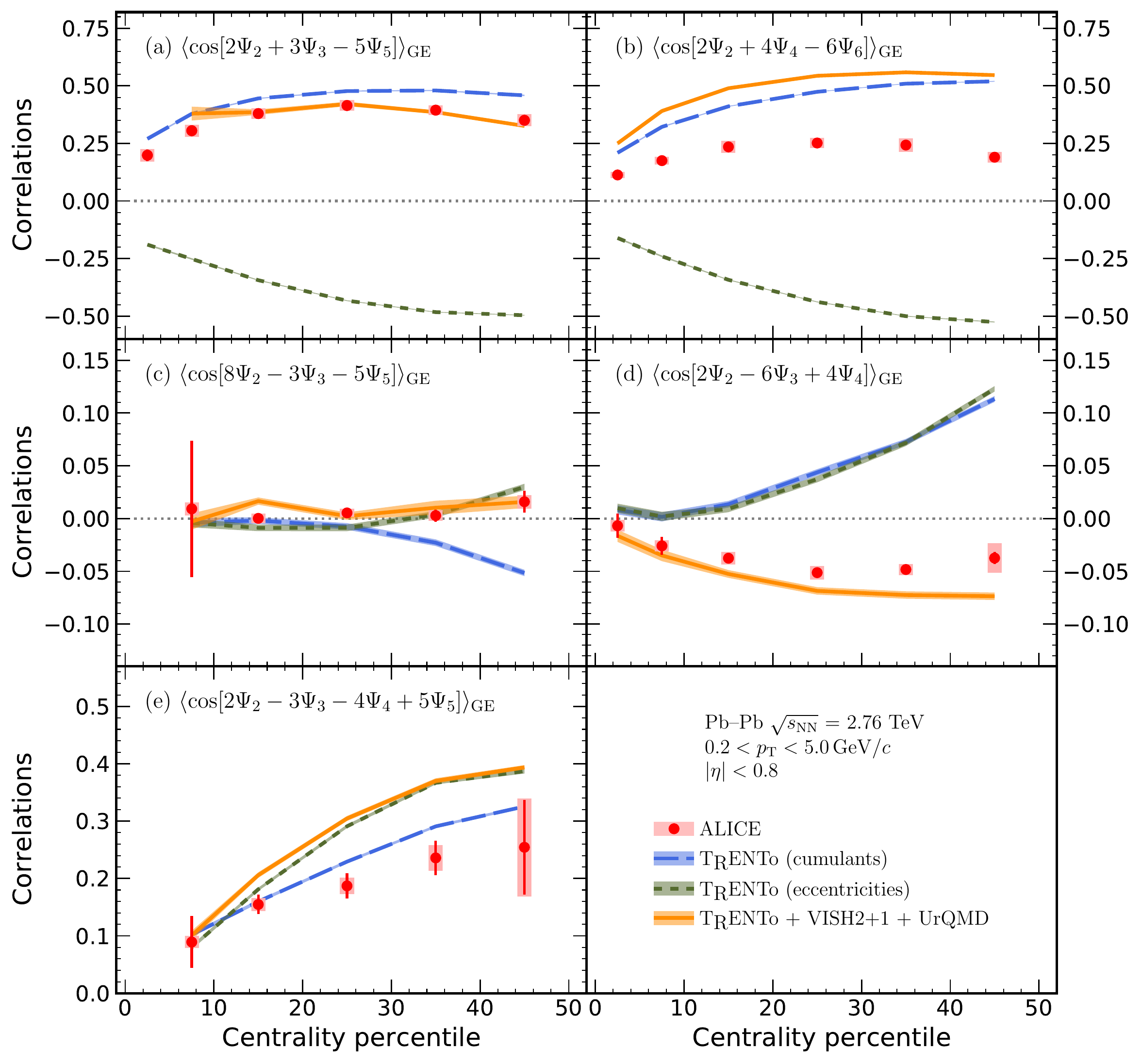}
	\caption{Experimental data for correlations between three (a-d) and four (e) symmetry planes compared with theoretical predictions in the initial and final state obtained with \trento\:and \trento+VISH(2+1)+UrQMD~\cite{Moreland:2014oya,Shen:2014vra,Song:2007ux,Bass:1998ca,Bleicher:1999xi}, respectively. Statistical (systematic) uncertainties of the ALICE data are shown as lines (boxes). The statistical uncertainties of the models are represented by the colored bands.} 
	\label{Fig3_TwoHar_models}
\end{figure}

The new measurements of SPC with the GE are compared with Monte Carlo simulations with the \trento+VISH(2+1)+UrQMD event generator~\cite{Moreland:2014oya,Shen:2014vra,Song:2007ux,Bass:1998ca,Bleicher:1999xi}. In this article, the maximum \textit{a posteriori} (MAP) estimation obtained in the Bayesian analysis in Ref.~\cite{Bernhard:2019bmu} is used for the parameters of the model. In inferring the MAP parameterization, a series of ALICE measurements (two- and four-particle correlations, charged particle multiplicities, etc.) were used as inputs into the Bayesian analysis, while the SPC are not included in these studies. Including new observables (e.g. SPC) in the Bayesian analysis can lead to an improvement in the uncertainty of the inferred parameter and resolving the discrepancies~\cite{Parkkila:2021tqq,Parkkila:2021yha}. If the discrepancy between model and data persists even after including new observables as input, the model itself needs to be revised. 

In addition to the model predictions of final-state SPC, initial-state participant plane correlations are studied with \trento. The participant plane of order $n$ takes the same role in the initial state as the symmetry plane in the final state. The correlations between participant planes are extracted from the initial state where flow vectors $v_ne^{in\Psi_n}$ are replaced first by eccentricities~\cite{Gardim:2011xv}, and second by cumulants of the initial energy density~\cite{Teaney:2010vd,Teaney:2012ke,Teaney:2013dta}.
The eccentricities are defined as
\begin{equation}
    \epsilon_n e^{in\phi_n} = -\frac{\{r^n e^{in\varphi}\}}{\{r^n\}},\qquad n>1\,,
\end{equation}
where $\{\cdots\}=\int r\text{d}r\text{d}\varphi\,\varepsilon(r,\varphi)$ stands for the average with respect to the initial energy density $\varepsilon(r,\varphi)$ in the transverse direction and $(r,\varphi)$ are the polar coordinates in the transverse plane. Eccentricities are the moments of the initial energy density distribution. The cumulants of the initial energy density distribution, $c_n e^{in\Phi_n}$, are obtained as a combination of eccentricities and the radial moments of the energy density, $\{r^n\}$. In fact, cumulants are a better measure to study the deformation of a distribution close to a Gaussian. Borrowing a motivating example from Ref.~\cite{Teaney:2012ke}, a Gaussian distribution $e^{-x^2/2\sigma_x^2-y^2/2\sigma_y^2}$ has infinitely many non-vanishing moments, while only its second order cumulants are non-zero.  Following the convention of Ref.~\cite{Teaney:2012ke}, the first two cumulants and eccentricities are equivalent, $c_n e^{in\Phi_n}= \epsilon_n e^{in\phi_n}$ for $n=2,3$. Higher order cumulants have non-trivial relations to eccentricities. Here, only the fourth harmonic is shown as an example:
\begin{equation}\label{n4InitialCumulant}
    c_4 e^{i4\Phi_4}=\epsilon_4 e^{i4\phi_4}+3\left(\frac{\{r^2\}^2}{\{r^4\}}\right)\epsilon_2^2 e^{i4\phi_2}.
\end{equation}
More details can be found in Refs.~\cite{Teaney:2010vd,Teaney:2012ke,Teaney:2013dta}.

The comparison with initial and final state SPC demonstrates how much of the observed correlation is inherited from the initial state. This is due to the linear and non-linear response of the medium~\cite{Yan:2017ivm}. For the second- and third-order anisotropies, the linear response is expected to dominate i.e. $v_2e^{i2\Psi_2} = \omega_2 c_2e^{i2\Phi_2}$ and $v_3e^{i3\Psi_3} = \omega_3 c_3e^{i3\Phi_3}$, especially in central and semicentral collisions~\cite{Gardim:2011xv,Niemi:2012aj,Noronha-Hostler:2015dbi}. The $\omega_i$ describe the linear hydrodynamic coupling constants. For higher orders, non-linear contributions will play a significant role, e.g. in case of the fourth order as 
\begin{equation}
    v_4e^{i4\Psi_4} = \omega_4 c_4e^{i4\Phi_4} + \omega_{422}c_2^2e^{i4\Phi_2}\,,
\end{equation}
where $\omega_{422}$ is the non-linear coupling between the second- and fourth-order anisotropies~\cite{Teaney:2010vd,Teaney:2012ke,Teaney:2013dta}. As an example of how this impacts the SPC, one can build the quantity $v_2^2v_4e^{i4(\Psi_4-\Psi_2)}$. The real part of its phase corresponds to the SPC $\SPCTwoFour$. Using the linear and non-linear response, one can translate this into the initial state as: 
\begin{equation}
v_2^2v_4e^{i4(\Psi_4-\Psi_2)} = \omega_2 \omega_4 c_2^2 c_4  e^{i4\left(\Phi_4-\Phi_2\right)} +  \omega_{422} \omega_2^2 c_2^2 \,.
\end{equation}
The latter equation shows that the initial and final state SPC $\SPCTwoFour$ will be equal to each other in the limit of pure linear response, while they will deviate in case of a non-zero non-linear coupling $\omega_{422}$.

Figure~\ref{Fig2_TwoHar_models} shows the comparison between the model calculations and the experimental data for the correlation between two symmetry planes. For the SPC $\TwoFour$ the initial-state participant plane correlations given by the energy density cumulants overlap with the final-state prediction of the SPC up to 10\% in centrality, indicating a vanishing non-linear coupling in the regime dominated by flow fluctuations. For higher centralities, the two curves increasingly deviate, showing the presence of a non-zero non-linear coupling between the second- and fourth-order flow vectors. In particular, it can be observed that the final-state prediction increasingly deviates from the data with increasing centrality. This is expected, since in this regime strong correlations between the second and fourth harmonics can originate from the initial ellipsoidal geometry. The non-linear coupling constant between initial-state ellipticity and $v_4 e^{i 4\Psi_4}$ from \trento~to iEBE-VISHNU was studied in Ref.~\cite{Taghavi:2020gcy}. It was demonstrated that this coupling is very small up to 10\% and it is positive up to 70\% centrality. For $\TwoThree$ the experimental data show a flat centrality behavior and are compatible with zero within the uncertainties. The model predicts small values for $\TwoThree$ with a flat centrality behavior although the predictions for the final state slightly overestimate the data. However, the initial-state correlations decrease to more negative values for centralities above 30\%. For $\TwoThree$, a linear response is expected to be an accurate approximation for harmonics $n=2, 3$ in central and semicentral collisions~\cite{Gardim:2011xv,Niemi:2012aj,Noronha-Hostler:2015dbi}. As such, one would expect that the initial state participant plane correlations should be the same as the final state symmetry plane correlations between the second- and third-order harmonics. One possible explanation is that higher-order terms beyond linear response are responsible for decreasing the correlation during the hydrodynamic evolution, and the final value accidentally lands on very small numbers. In this respect, more rigorous study is needed in the future. While the model captures the qualitative behavior of the experimental signal in the two previous cases, a quantitative agreement is not observed in every case, particularly not for SPC involving $\Psi_6$ or correlations between four symmetry planes. For the SPC $\TwoSix$ and $\ThreeSix$, a large deviation between the data and the final-state model prediction can be observed. This deviation could be related to the complex dynamics of the sixth-order symmetry plane, which involves multiple non-linear responses to the lower order symmetry planes $\Psi_2$, $\Psi_3$ and $\Psi_4$. For the SPC $\ThreeSix$ it is in particular interesting that the initial-state correlation becomes stronger with increasing centrality, while the final-state correlation becomes weaker with increasing centrality in the model. In contrast to that, the experimental data shows only a very weak centrality dependence. 

The sign change between correlations obtained from eccentricities (green short-dashed curves) and initial energy density cumulants (blue long-dashed curves) in Fig.~\ref{Fig2_TwoHar_models} was pointed out in Ref.~\cite{Teaney:2013dta}. The actual shape of the initial state is captured by $c_ne^{in\Phi_n}$. As a result, the linear hydrodynamic response approximation is more accurate when employing cumulants. As an example, SPC $\TwoFour$ in Fig.~\ref{Fig2_TwoHar_models} panel (a) is considered. Referring to Eq.~\eqref{n4InitialCumulant}, the difference between $\epsilon_4 e^{in\phi_4}$ and $c_4 e^{in\Phi_4}$ is proportional to $\{r^2\}^2/\{r^4\}$, which merely depends on the radial shape of the initial state. The numerator of the correlation $\langle \cos (4\phi_2-4\phi_4)\rangle_{\text{GE}}$ is proportional to the real part of $\langle\epsilon_2^2\epsilon_4 e^{4i\phi_2-4i\phi_4}\rangle$. Substituting from Eq.~\eqref{n4InitialCumulant} ($c_2e^{2i\Phi_2}$ and $\epsilon_2e^{2i\phi_2}$ are identical), one finds $\langle\epsilon_2^2\epsilon_4 e^{4i\phi_2-4i\phi_4}\rangle = \langle c_2^2c_4 e^{4i\Phi_2-4i\Phi_4}\rangle  -3\langle (\{r^2\}^2/\{r^4\}) c_2^4\rangle $. In case the second term on the right is small, SPC of cumulants are expected to be close to those calculated from eccentricities. However, noting that $3\langle (\{r^2\}^2/\{r^4\}) c_2^4\rangle $ is real and positive, the contribution of this term is such that it leads to a negative sign for $\langle\epsilon_2^2\epsilon_4 e^{4i\phi_2-4i\phi_4}\rangle$. Therefore, one concludes that, regarding the hydrodynamic response, using energy-density cumulants for the initial state is more appropriate.  A more detailed study can be found in Ref.~\cite{Taghavi:2020gcy}.

In Fig.~\ref{Fig3_TwoHar_models}, the same comparison of model predictions with respect to the experimental data was performed for SPC involving three or four symmetry planes. For the SPC $\TwoThreeFive$, a non-zero correlation signal was extracted, which existed to a larger extent already in the initial state. The final-state predictions of \trento+iEBE-VISHNU are in good agreement with the experimental data. Further considering the same three symmetry planes $\Psi_2$, $\Psi_3$ and $\Psi_5$, the combination $\SecondTwoThreeFive$ results in an experimental signal that is compatible with zero within uncertainties for all reported centrality intervals. The deviation between $\TwoThreeFive$ and $\SecondTwoThreeFive$ can be attributed to different contributions from the initial-state in the non-linear response of the two observables. While the non-linear response term for $\TwoThreeFive$ does not contain contributions from any participant plane correlations, the non-linear part of $\SecondTwoThreeFive$ picks up such an additional correlation from the initial state. In particular, the SPC $\SecondTwoThreeFive$ has a non-linear coupling to the initial-state correlation between the second- and third-order participant planes. Similar to previously presented examples of SPC that involve the sixth-order symmetry plane $\Psi_6$, the final-state model prediction for $\TwoFourSix$ shows a large deviation compared to the measurements. This deviation is again attributed to the complexity of $\Psi_6$, which makes it particularly sensitive to the model parameters. One finds that $\TwoThreeFive$ has the strongest signal in data while $\TwoFourSix$ is larger than $\TwoThreeFive$ in the model. The results of $\TwoThreeFour$ are the only measurement with a negative signal for the SPC. While the model predicts this behavior as well, it can be observed that the initial-state correlations are strictly non-negative. Thus, the sign-change between the initial and final state can be linked to the hydrodynamic evolution of the system. Lastly, Fig.~\ref{Fig3_TwoHar_models} shows the comparison with the first experimental measurement of correlations between four symmetry planes. While the final-state model prediction captures the qualitative behavior, a quantitative agreement between experimental data and model is not observed. A recent Bayesian analysis of the QGP hydrodynamic properties~\cite{Parkkila:2021yha} has shown that higher-order correlations as well as measurements involving higher order flow harmonics are more sensitive to changes in the QGP parameters. Thus, in particular the measurements of SPC including $\Psi_6$ and the correlations between four symmetry planes are expected to give more stringent constraints in future analyses of this kind.

\section{Conclusion}
\label{s:Conclusion}
Utilizing the recently introduced Gaussian Estimator, the first measurements of symmetry plane correlations, which are not influenced by correlations between different flow amplitudes, are presented in Pb--Pb collisions at $\sqrt{s_\text{NN}}=$~2.76~TeV. Correlations between two, three and, for the first time, four symmetry planes were shown. The data show a clear order for the different SPC, which can be related to the cumulants of the underlying flow vector fluctuations. The measurements using the new GE show significantly smaller symmetry plane correlations than previously reported by the SP method. This observation is in qualitative agreement with the study in Ref.~\cite{Bilandzic:2020csw} which reported the bias of the SP method to larger values due to the influence of correlations between the flow amplitudes. In contrast to the SP method, the results of $\TwoThree$ are consistent with zero with the current uncertainties. Within the uncertainties, this shows that $\Psi_2$ and $\Psi_3$ are fully uncorrelated, which was qualitatively reported by a previous ALICE study~\cite{ALICE:2011ab}. Future studies using Run 2 Pb--Pb data as well as the upcoming Run 3 Pb--Pb campaign will show whether a small non-zero correlation exists between $\Psi_2$ and $\Psi_3$, which cannot be resolved within the present uncertainties. Employing state-of-the-art hydrodynamic model calculations, one could see that the predictions and the measurements are not in quantitative agreement for all SPC. The most significant deviations are observed in the correlations $\ThreeSix$, $\TwoSix$, $\TwoFourSix$ and $\TwoThreeFourFive$. Future studies have to address how the initial state correlations between the second- and third-order participant planes are suppressed in the final state SPC $\TwoThree$ as one would expect linear response to be a good approximation for the second- and third-order flow harmonics. Since the measured SPC contain independent information about flow vector fluctuations, they will provide useful inputs for future Bayesian analyses aiming at extracting the properties of the QGP.


\newenvironment{acknowledgement}{\relax}{\relax}
\begin{acknowledgement}
\section*{Acknowledgements}

The ALICE Collaboration would like to thank all its engineers and technicians for their invaluable contributions to the construction of the experiment and the CERN accelerator teams for the outstanding performance of the LHC complex.
The ALICE Collaboration gratefully acknowledges the resources and support provided by all Grid centres and the Worldwide LHC Computing Grid (WLCG) collaboration.
The ALICE Collaboration acknowledges the following funding agencies for their support in building and running the ALICE detector:
A. I. Alikhanyan National Science Laboratory (Yerevan Physics Institute) Foundation (ANSL), State Committee of Science and World Federation of Scientists (WFS), Armenia;
Austrian Academy of Sciences, Austrian Science Fund (FWF): [M 2467-N36] and Nationalstiftung f\"{u}r Forschung, Technologie und Entwicklung, Austria;
Ministry of Communications and High Technologies, National Nuclear Research Center, Azerbaijan;
Conselho Nacional de Desenvolvimento Cient\'{\i}fico e Tecnol\'{o}gico (CNPq), Financiadora de Estudos e Projetos (Finep), Funda\c{c}\~{a}o de Amparo \`{a} Pesquisa do Estado de S\~{a}o Paulo (FAPESP) and Universidade Federal do Rio Grande do Sul (UFRGS), Brazil;
Bulgarian Ministry of Education and Science, within the National Roadmap for Research Infrastructures 2020-2027 (object CERN), Bulgaria;
Ministry of Education of China (MOEC) , Ministry of Science \& Technology of China (MSTC) and National Natural Science Foundation of China (NSFC), China;
Ministry of Science and Education and Croatian Science Foundation, Croatia;
Centro de Aplicaciones Tecnol\'{o}gicas y Desarrollo Nuclear (CEADEN), Cubaenerg\'{\i}a, Cuba;
Ministry of Education, Youth and Sports of the Czech Republic, Czech Republic;
The Danish Council for Independent Research | Natural Sciences, the VILLUM FONDEN and Danish National Research Foundation (DNRF), Denmark;
Helsinki Institute of Physics (HIP), Finland;
Commissariat \`{a} l'Energie Atomique (CEA) and Institut National de Physique Nucl\'{e}aire et de Physique des Particules (IN2P3) and Centre National de la Recherche Scientifique (CNRS), France;
Bundesministerium f\"{u}r Bildung und Forschung (BMBF) and GSI Helmholtzzentrum f\"{u}r Schwerionenforschung GmbH, Germany;
General Secretariat for Research and Technology, Ministry of Education, Research and Religions, Greece;
National Research, Development and Innovation Office, Hungary;
Department of Atomic Energy Government of India (DAE), Department of Science and Technology, Government of India (DST), University Grants Commission, Government of India (UGC) and Council of Scientific and Industrial Research (CSIR), India;
National Research and Innovation Agency - BRIN, Indonesia;
Istituto Nazionale di Fisica Nucleare (INFN), Italy;
Japanese Ministry of Education, Culture, Sports, Science and Technology (MEXT) and Japan Society for the Promotion of Science (JSPS) KAKENHI, Japan;
Consejo Nacional de Ciencia (CONACYT) y Tecnolog\'{i}a, through Fondo de Cooperaci\'{o}n Internacional en Ciencia y Tecnolog\'{i}a (FONCICYT) and Direcci\'{o}n General de Asuntos del Personal Academico (DGAPA), Mexico;
Nederlandse Organisatie voor Wetenschappelijk Onderzoek (NWO), Netherlands;
The Research Council of Norway, Norway;
Commission on Science and Technology for Sustainable Development in the South (COMSATS), Pakistan;
Pontificia Universidad Cat\'{o}lica del Per\'{u}, Peru;
Ministry of Education and Science, National Science Centre and WUT ID-UB, Poland;
Korea Institute of Science and Technology Information and National Research Foundation of Korea (NRF), Republic of Korea;
Ministry of Education and Scientific Research, Institute of Atomic Physics, Ministry of Research and Innovation and Institute of Atomic Physics and University Politehnica of Bucharest, Romania;
Ministry of Education, Science, Research and Sport of the Slovak Republic, Slovakia;
National Research Foundation of South Africa, South Africa;
Swedish Research Council (VR) and Knut \& Alice Wallenberg Foundation (KAW), Sweden;
European Organization for Nuclear Research, Switzerland;
Suranaree University of Technology (SUT), National Science and Technology Development Agency (NSTDA), Thailand Science Research and Innovation (TSRI) and National Science, Research and Innovation Fund (NSRF), Thailand;
Turkish Energy, Nuclear and Mineral Research Agency (TENMAK), Turkey;
National Academy of  Sciences of Ukraine, Ukraine;
Science and Technology Facilities Council (STFC), United Kingdom;
National Science Foundation of the United States of America (NSF) and United States Department of Energy, Office of Nuclear Physics (DOE NP), United States of America.
In addition, individual groups or members have received support from:
European Research Council, Strong 2020 - Horizon 2020, Marie Sk\l{}odowska Curie (grant nos. 950692, 824093, 896850), European Union;
Academy of Finland (Center of Excellence in Quark Matter) (grant nos. 346327, 346328), Finland;
Programa de Apoyos para la Superaci\'{o}n del Personal Acad\'{e}mico, UNAM, Mexico.

\end{acknowledgement}

\bibliographystyle{utphys}   
\nocite{*}
\bibliography{bibliography.bib}

\newpage
\appendix
\section{A brief overview of the Gaussian Estimator \label{Sec. GE-Example}} 

The details of the derivation and validation of the GE in Eq.~\eqref{GaussApproxExp} can be found in Ref.~\cite{Bilandzic:2020csw}. In the following, a brief overview of the main idea is presented. The concept of the GE is similar to estimating the average elliptic flow originating from the geometry of the collision by using the flow measurements $v_2\{2k\}$. To briefly explain, let us define $v_{2,x}=v_2\cos2(\Psi_2-\Psi_{\text{RP}})$ and $v_{2,y}=v_2\sin2(\Psi_2-\Psi_{\text{RP}})$. Since $\Psi_2$ fluctuates around the reaction plane angle $\Psi_{\text{RP}}$, the average $\overline{v}_2 = \langle v_{2,x} \rangle $ depends merely on the nonzero average of the initial elliptic shape and $\langle v_{2,y} \rangle$ vanishes~\cite{Voloshin:2007pc}. However, the angle $\Psi_{\text{RP}}$ rotates randomly event-by-event in the experiment, therefore, $\overline{v}_2$ is not accessible directly. To estimate the value of $\overline{v}_2$, one employs the central limit theorem to approximate the fluctuation of $(v_{2,x},v_{2,y})$ as a 2D Gaussian proportional to $\exp\left[-((v_{2,x}-\overline{v}_2)^2+v_{2,y}^2)/2\sigma_v^2\right]=\exp\left[-(v_2^2+\overline{v}_2^2-2v_2\overline{v}_2\cos2\Psi_2)/2\sigma_v^2\right]$. By averaging over $\Psi_2$, one finds a Bessel--Gaussian distribution depending on the random variable $v_2$ where $\overline{v}_2$ is a parameter that controls the shape of the distribution~\cite{Voloshin:1994mz,E877:1994plr,Voloshin:2007pc}. Using this information, one finds the estimation $\overline{v}_2 \approx v_2\{2k\}$ for $k>1$~\cite{Voloshin:2007pc}.

To explain the idea behind Eq.~\eqref{GaussApproxExp}, the SPC $\langle\cos 4(\Psi_4-\Psi_2)\rangle$ is considered as a simple. More general cases can be obtained accordingly. In order to estimate $\langle\cos 4(\Psi_4-\Psi_2)\rangle$ in terms of  quantities such as $\langle v_2^2v_4 \cos 4(\Psi_4-\Psi_2)\rangle$, the two variables $R=v_2^2v_4$ and $\theta=4(\Psi_4-\Psi_2)$ are defined, or in Cartesian form $R_x=R\cos\theta$ and $R_y=R\sin\theta$. Since $R$ and $\theta$ are correlated, the ratio $\langle R \cos\theta \rangle /\langle R \rangle $ would not be equal to $\langle\cos\theta\rangle$. However, the $(R_x,R_y)$ fluctuation can be approximated with a 2D Gaussian distribution,
\begin{equation}\label{2DGauss}
N(R_x,R_y)=\frac{1}{\pi \sigma_R^2}\exp\left[-\frac{(R_x-\mu_x)^2+R_y^2}{\sigma_R^2}\right],
\end{equation}
where 
\begin{equation}\label{2Dparams}
\mu_x = \langle R_x \rangle = \langle v_2^2 v_4 \cos  4(\Psi_4-\Psi_2)  \rangle,\qquad \sigma_R\approx \sqrt{\langle R_x^2\rangle + \langle R_y^2\rangle} =\sqrt{\langle v_2^4 v_4^2 \rangle}.
\end{equation}
The equation \eqref{2DGauss} in polar coordinates is proportional to $\exp[-(R^2+\mu_x^2-2R\mu_x\cos\theta)/\sigma_R^2]$. Averaging over the variable $\theta$ would lead to a Bessel-Gaussian distribution. However, the goal of this study is the extraction of information about $\cos\theta$ fluctuations. Therefore, one needs to average out the variable $R$ to find a distribution that depends on $\theta$ only. As a result, one finds
\begin{equation}\label{cosMoment}
	\langle \cos\theta\rangle \approx \sqrt{\frac{\pi}{4}}\left(\frac{\mu_x}{\sigma_R}\right),
\end{equation}
where the prefactor $\sqrt{\pi/4} \approx 0.886$ is the consequence of integration over $R$ and $\theta$ in the calculation of $\langle \cos\theta\rangle$.  
Substituting Eq.~\eqref{2Dparams} into Eq.~\eqref{cosMoment}, one finds
\begin{equation}
\langle\cos 4(\Psi_4-\Psi_2)\rangle_{\text{GE}} = \sqrt{\frac{\pi}{4}} \;\frac{\langle v_2^2 v_4 \cos  4(\Psi_4-\Psi_2)  \rangle}{\sqrt{\langle v_2^4 v_4^2 \rangle}},
\end{equation}
which is a special case of Eq.~\eqref{GaussApproxExp}. To derive the analytical expression in Eq.~\eqref{cosMoment}, an expansion up to leading terms with respect to $\mu_x/\sigma_R$ was considered. Also it has been assumed that $\langle R_x^2\rangle \approx \langle R_y^2\rangle$. On top of these assumptions, the $(R_x,R_y)$ fluctuation should be close to a 2D Gaussian to have an accurate estimation. Comparing with the true values of the SPC in the hydrodynamic simulations, it turns out that these approximations lead to an accurate estimation~\cite{Bilandzic:2020csw}.  In the scalar product method, the correlation is given by
\begin{equation}
\langle\cos 4(\Psi_4-\Psi_2)\rangle_{\text{SP}} = \frac{\langle v_2^2 v_4 \cos  4(\Psi_4-\Psi_2)  \rangle}{\sqrt{\langle v_2^4\rangle \langle v_4^2 \rangle}}.
\end{equation}
Apart from the numerical prefactor, the denominator of GE contains a joint correlation $\langle v_2^4 v_4^2 \rangle$ rather than $\langle v_2^4 \rangle \langle v_4^2 \rangle$ as in the SP method. 


\newpage

\section{The ALICE Collaboration}
\label{app:collab}
\begin{flushleft} 
\small

S.~Acharya\,\orcidlink{0000-0002-9213-5329}\,$^{\rm 125}$, 
D.~Adamov\'{a}\,\orcidlink{0000-0002-0504-7428}\,$^{\rm 86}$, 
A.~Adler$^{\rm 69}$, 
G.~Aglieri Rinella\,\orcidlink{0000-0002-9611-3696}\,$^{\rm 32}$, 
M.~Agnello\,\orcidlink{0000-0002-0760-5075}\,$^{\rm 29}$, 
N.~Agrawal\,\orcidlink{0000-0003-0348-9836}\,$^{\rm 50}$, 
Z.~Ahammed\,\orcidlink{0000-0001-5241-7412}\,$^{\rm 132}$, 
S.~Ahmad\,\orcidlink{0000-0003-0497-5705}\,$^{\rm 15}$, 
S.U.~Ahn\,\orcidlink{0000-0001-8847-489X}\,$^{\rm 70}$, 
I.~Ahuja\,\orcidlink{0000-0002-4417-1392}\,$^{\rm 37}$
% ...

\section*{Affiliation Notes}

$^{\rm I}$ Deceased\\
$^{\rm II}$ Also at: Max-Planck-Institut f\"{u}r Physik, Munich, Germany\\
$^{\rm III}$ Also at: Italian National Agency for New Technologies, Energy and Sustainable Economic Development (ENEA), Bologna, Italy\\
$^{\rm IV}$ Also at: Department of Applied Physics, Aligarh Muslim University, Aligarh, India\\
$^{\rm V}$ Also at: Institute of Theoretical Physics, University of Wroclaw, Poland\\
$^{\rm VI}$ Also at: An institution covered by a cooperation agreement with CERN\\

\section*{Collaboration Institutes}

$^{1}$ A.I. Alikhanyan National Science Laboratory (Yerevan Physics Institute) Foundation, Yerevan, Armenia\\
$^{2}$ AGH University of Science and Technology, Cracow, Poland\\
$^{3}$ Bogolyubov Institute for Theoretical Physics, National Academy of Sciences of Ukraine, Kiev, Ukraine\\
$^{4}$ Bose Institute, Department of Physics  and Centre for Astroparticle Physics and Space Science (CAPSS), Kolkata, India\\
$^{5}$ California Polytechnic State University, San Luis Obispo, California, United States\\
$^{6}$ Central China Normal University, Wuhan, China\\
$^{7}$ Centro de Aplicaciones Tecnol\'{o}gicas y Desarrollo Nuclear (CEADEN), Havana, Cuba\\
$^{8}$ Centro de Investigaci\'{o}n y de Estudios Avanzados (CINVESTAV), Mexico City and M\'{e}rida, Mexico\\
$^{9}$ Chicago State University, Chicago, Illinois, United States\\
$^{10}$ China Institute of Atomic Energy, Beijing, China\\
$^{11}$ Chungbuk National University, Cheongju, Republic of Korea\\
$^{12}$ Comenius University Bratislava, Faculty of Mathematics, Physics and Informatics, Bratislava, Slovak Republic\\
$^{13}$ COMSATS University Islamabad, Islamabad, Pakistan\\
$^{14}$ Creighton University, Omaha, Nebraska, United States\\
$^{15}$ Department of Physics, Aligarh Muslim University, Aligarh, India\\
$^{16}$ Department of Physics, Pusan National University, Pusan, Republic of Korea\\
$^{17}$ Department of Physics, Sejong University, Seoul, Republic of Korea\\
$^{18}$ Department of Physics, University of California, Berkeley, California, United States\\
$^{19}$ Department of Physics, University of Oslo, Oslo, Norway\\
$^{20}$ Department of Physics and Technology, University of Bergen, Bergen, Norway\\
$^{21}$ Dipartimento di Fisica, Universit\`{a} di Pavia, Pavia, Italy\\
$^{22}$ Dipartimento di Fisica dell'Universit\`{a} and Sezione INFN, Cagliari, Italy\\
$^{23}$ Dipartimento di Fisica dell'Universit\`{a} and Sezione INFN, Trieste, Italy\\
$^{24}$ Dipartimento di Fisica dell'Universit\`{a} and Sezione INFN, Turin, Italy\\
$^{25}$ Dipartimento di Fisica e Astronomia dell'Universit\`{a} and Sezione INFN, Bologna, Italy\\
$^{26}$ Dipartimento di Fisica e Astronomia dell'Universit\`{a} and Sezione INFN, Catania, Italy\\
$^{27}$ Dipartimento di Fisica e Astronomia dell'Universit\`{a} and Sezione INFN, Padova, Italy\\
$^{28}$ Dipartimento di Fisica `E.R.~Caianiello' dell'Universit\`{a} and Gruppo Collegato INFN, Salerno, Italy\\
$^{29}$ Dipartimento DISAT del Politecnico and Sezione INFN, Turin, Italy\\
$^{30}$ Dipartimento di Scienze MIFT, Universit\`{a} di Messina, Messina, Italy\\
$^{31}$ Dipartimento Interateneo di Fisica `M.~Merlin' and Sezione INFN, Bari, Italy\\
$^{32}$ European Organization for Nuclear Research (CERN), Geneva, Switzerland\\
$^{33}$ Faculty of Electrical Engineering, Mechanical Engineering and Naval Architecture, University of Split, Split, Croatia\\
$^{34}$ Faculty of Engineering and Science, Western Norway University of Applied Sciences, Bergen, Norway\\
$^{35}$ Faculty of Nuclear Sciences and Physical Engineering, Czech Technical University in Prague, Prague, Czech Republic\\
$^{36}$ Faculty of Physics, Sofia University, Sofia, Bulgaria\\
$^{37}$ Faculty of Science, P.J.~\v{S}af\'{a}rik University, Ko\v{s}ice, Slovak Republic\\
$^{38}$ Frankfurt Institute for Advanced Studies, Johann Wolfgang Goethe-Universit\"{a}t Frankfurt, Frankfurt, Germany\\
$^{39}$ Fudan University, Shanghai, China\\
$^{40}$ Gangneung-Wonju National University, Gangneung, Republic of Korea\\
$^{41}$ Gauhati University, Department of Physics, Guwahati, India\\
$^{42}$ Helmholtz-Institut f\"{u}r Strahlen- und Kernphysik, Rheinische Friedrich-Wilhelms-Universit\"{a}t Bonn, Bonn, Germany\\
$^{43}$ Helsinki Institute of Physics (HIP), Helsinki, Finland\\
$^{44}$ High Energy Physics Group,  Universidad Aut\'{o}noma de Puebla, Puebla, Mexico\\
$^{45}$ Horia Hulubei National Institute of Physics and Nuclear Engineering, Bucharest, Romania\\
$^{46}$ Indian Institute of Technology Bombay (IIT), Mumbai, India\\
$^{47}$ Indian Institute of Technology Indore, Indore, India\\
$^{48}$ INFN, Laboratori Nazionali di Frascati, Frascati, Italy\\
$^{49}$ INFN, Sezione di Bari, Bari, Italy\\
$^{50}$ INFN, Sezione di Bologna, Bologna, Italy\\
$^{51}$ INFN, Sezione di Cagliari, Cagliari, Italy\\
$^{52}$ INFN, Sezione di Catania, Catania, Italy\\
$^{53}$ INFN, Sezione di Padova, Padova, Italy\\
$^{54}$ INFN, Sezione di Pavia, Pavia, Italy\\
$^{55}$ INFN, Sezione di Torino, Turin, Italy\\
$^{56}$ INFN, Sezione di Trieste, Trieste, Italy\\
$^{57}$ Inha University, Incheon, Republic of Korea\\
$^{58}$ Institute for Gravitational and Subatomic Physics (GRASP), Utrecht University/Nikhef, Utrecht, Netherlands\\
$^{59}$ Institute of Experimental Physics, Slovak Academy of Sciences, Ko\v{s}ice, Slovak Republic\\
$^{60}$ Institute of Physics, Homi Bhabha National Institute, Bhubaneswar, India\\
$^{61}$ Institute of Physics of the Czech Academy of Sciences, Prague, Czech Republic\\
$^{62}$ Institute of Space Science (ISS), Bucharest, Romania\\
$^{63}$ Institut f\"{u}r Kernphysik, Johann Wolfgang Goethe-Universit\"{a}t Frankfurt, Frankfurt, Germany\\
$^{64}$ Instituto de Ciencias Nucleares, Universidad Nacional Aut\'{o}noma de M\'{e}xico, Mexico City, Mexico\\
$^{65}$ Instituto de F\'{i}sica, Universidade Federal do Rio Grande do Sul (UFRGS), Porto Alegre, Brazil\\
$^{66}$ Instituto de F\'{\i}sica, Universidad Nacional Aut\'{o}noma de M\'{e}xico, Mexico City, Mexico\\
$^{67}$ iThemba LABS, National Research Foundation, Somerset West, South Africa\\
$^{68}$ Jeonbuk National University, Jeonju, Republic of Korea\\
$^{69}$ Johann-Wolfgang-Goethe Universit\"{a}t Frankfurt Institut f\"{u}r Informatik, Fachbereich Informatik und Mathematik, Frankfurt, Germany\\
$^{70}$ Korea Institute of Science and Technology Information, Daejeon, Republic of Korea\\
$^{71}$ KTO Karatay University, Konya, Turkey\\
$^{72}$ Laboratoire de Physique des 2 Infinis, Ir\`{e}ne Joliot-Curie, Orsay, France\\
$^{73}$ Laboratoire de Physique Subatomique et de Cosmologie, Universit\'{e} Grenoble-Alpes, CNRS-IN2P3, Grenoble, France\\
$^{74}$ Lawrence Berkeley National Laboratory, Berkeley, California, United States\\
$^{75}$ Lund University Department of Physics, Division of Particle Physics, Lund, Sweden\\
$^{76}$ Nagasaki Institute of Applied Science, Nagasaki, Japan\\
$^{77}$ Nara Women{'}s University (NWU), Nara, Japan\\
$^{78}$ National and Kapodistrian University of Athens, School of Science, Department of Physics , Athens, Greece\\
$^{79}$ National Centre for Nuclear Research, Warsaw, Poland\\
$^{80}$ National Institute of Science Education and Research, Homi Bhabha National Institute, Jatni, India\\
$^{81}$ National Nuclear Research Center, Baku, Azerbaijan\\
$^{82}$ National Research and Innovation Agency - BRIN, Jakarta, Indonesia\\
$^{83}$ Niels Bohr Institute, University of Copenhagen, Copenhagen, Denmark\\
$^{84}$ Nikhef, National institute for subatomic physics, Amsterdam, Netherlands\\
$^{85}$ Nuclear Physics Group, STFC Daresbury Laboratory, Daresbury, United Kingdom\\
$^{86}$ Nuclear Physics Institute of the Czech Academy of Sciences, Husinec-\v{R}e\v{z}, Czech Republic\\
$^{87}$ Oak Ridge National Laboratory, Oak Ridge, Tennessee, United States\\
$^{88}$ Ohio State University, Columbus, Ohio, United States\\
$^{89}$ Physics department, Faculty of science, University of Zagreb, Zagreb, Croatia\\
$^{90}$ Physics Department, Panjab University, Chandigarh, India\\
$^{91}$ Physics Department, University of Jammu, Jammu, India\\
$^{92}$ Physics Program and International Institute for Sustainability with Knotted Chiral Meta Matter (SKCM2), Hiroshima University, Hiroshima, Japan\\
$^{93}$ Physikalisches Institut, Eberhard-Karls-Universit\"{a}t T\"{u}bingen, T\"{u}bingen, Germany\\
$^{94}$ Physikalisches Institut, Ruprecht-Karls-Universit\"{a}t Heidelberg, Heidelberg, Germany\\
$^{95}$ Physik Department, Technische Universit\"{a}t M\"{u}nchen, Munich, Germany\\
$^{96}$ Politecnico di Bari and Sezione INFN, Bari, Italy\\
$^{97}$ Research Division and ExtreMe Matter Institute EMMI, GSI Helmholtzzentrum f\"ur Schwerionenforschung GmbH, Darmstadt, Germany\\
$^{98}$ Saga University, Saga, Japan\\
$^{99}$ Saha Institute of Nuclear Physics, Homi Bhabha National Institute, Kolkata, India\\
$^{100}$ School of Physics and Astronomy, University of Birmingham, Birmingham, United Kingdom\\
$^{101}$ Secci\'{o}n F\'{\i}sica, Departamento de Ciencias, Pontificia Universidad Cat\'{o}lica del Per\'{u}, Lima, Peru\\
$^{102}$ Stefan Meyer Institut f\"{u}r Subatomare Physik (SMI), Vienna, Austria\\
$^{103}$ SUBATECH, IMT Atlantique, Nantes Universit\'{e}, CNRS-IN2P3, Nantes, France\\
$^{104}$ Sungkyunkwan University, Suwon City, Republic of Korea\\
$^{105}$ Suranaree University of Technology, Nakhon Ratchasima, Thailand\\
$^{106}$ Technical University of Ko\v{s}ice, Ko\v{s}ice, Slovak Republic\\
$^{107}$ The Henryk Niewodniczanski Institute of Nuclear Physics, Polish Academy of Sciences, Cracow, Poland\\
$^{108}$ The University of Texas at Austin, Austin, Texas, United States\\
$^{109}$ Universidad Aut\'{o}noma de Sinaloa, Culiac\'{a}n, Mexico\\
$^{110}$ Universidade de S\~{a}o Paulo (USP), S\~{a}o Paulo, Brazil\\
$^{111}$ Universidade Estadual de Campinas (UNICAMP), Campinas, Brazil\\
$^{112}$ Universidade Federal do ABC, Santo Andre, Brazil\\
$^{113}$ University of Cape Town, Cape Town, South Africa\\
$^{114}$ University of Houston, Houston, Texas, United States\\
$^{115}$ University of Jyv\"{a}skyl\"{a}, Jyv\"{a}skyl\"{a}, Finland\\
$^{116}$ University of Kansas, Lawrence, Kansas, United States\\
$^{117}$ University of Liverpool, Liverpool, United Kingdom\\
$^{118}$ University of Science and Technology of China, Hefei, China\\
$^{119}$ University of South-Eastern Norway, Kongsberg, Norway\\
$^{120}$ University of Tennessee, Knoxville, Tennessee, United States\\
$^{121}$ University of the Witwatersrand, Johannesburg, South Africa\\
$^{122}$ University of Tokyo, Tokyo, Japan\\
$^{123}$ University of Tsukuba, Tsukuba, Japan\\
$^{124}$ University Politehnica of Bucharest, Bucharest, Romania\\
$^{125}$ Universit\'{e} Clermont Auvergne, CNRS/IN2P3, LPC, Clermont-Ferrand, France\\
$^{126}$ Universit\'{e} de Lyon, CNRS/IN2P3, Institut de Physique des 2 Infinis de Lyon, Lyon, France\\
$^{127}$ Universit\'{e} de Strasbourg, CNRS, IPHC UMR 7178, F-67000 Strasbourg, France, Strasbourg, France\\
$^{128}$ Universit\'{e} Paris-Saclay Centre d'Etudes de Saclay (CEA), IRFU, D\'{e}partment de Physique Nucl\'{e}aire (DPhN), Saclay, France\\
$^{129}$ Universit\`{a} degli Studi di Foggia, Foggia, Italy\\
$^{130}$ Universit\`{a} del Piemonte Orientale, Vercelli, Italy\\
$^{131}$ Universit\`{a} di Brescia, Brescia, Italy\\
$^{132}$ Variable Energy Cyclotron Centre, Homi Bhabha National Institute, Kolkata, India\\
$^{133}$ Warsaw University of Technology, Warsaw, Poland\\
$^{134}$ Wayne State University, Detroit, Michigan, United States\\
$^{135}$ Westf\"{a}lische Wilhelms-Universit\"{a}t M\"{u}nster, Institut f\"{u}r Kernphysik, M\"{u}nster, Germany\\
$^{136}$ Wigner Research Centre for Physics, Budapest, Hungary\\
$^{137}$ Yale University, New Haven, Connecticut, United States\\
$^{138}$ Yonsei University, Seoul, Republic of Korea\\
$^{139}$  Zentrum  f\"{u}r Technologie und Transfer (ZTT), Worms, Germany\\
$^{140}$ Affiliated with an institute covered by a cooperation agreement with CERN\\
$^{141}$ Affiliated with an international laboratory covered by a cooperation agreement with CERN.\\

\end{flushleft} 

\end{document}